\documentclass[twocolumn]{autart}
\usepackage{graphicx}
\usepackage{xcolor}
\usepackage{amsmath,amsfonts}

\usepackage{subfiles}
\usepackage{subcaption}
\usepackage{enumitem}
\usepackage{todonotes}
\usepackage[numbers,sort&compress]{natbib}
\usepackage{tikz}
\usetikzlibrary{decorations.markings}
\usetikzlibrary{shapes,arrows,positioning}
\usetikzlibrary{arrows.meta}
\usepackage[siunitx,american voltages, american currents]{circuitikz}
\usepackage{mdframed}
\usetikzlibrary{matrix}

\renewcommand{\S}{\mathcal{S}}
\newcommand{\s}{\mathbb{S}}
\newcommand{\ns}{{n_s}}
\newcommand{\nf}{{n_f}}
\newcommand{\B}{\mathcal{B}}
\newcommand{\ve}{\varepsilon}
\newcommand{\R}{\mathbb{R}}
\newcommand{\N}{\mathbb{N}}
\newcommand{\parc}[1]{\frac{\partial }{\partial #1}}
\newcommand{\parcs}[2]{\frac{\partial #1}{\partial #2}}

\newcommand{\I}{\mathbb{I}}

\begin{document}

\begin{frontmatter}
	\title{Stabilization of slow-fast systems at fold points}
	\author[rug]{H. Jard\'on-Kojakhmetov}\ead{h.jardon.kojakhmetov@rug.nl},    
	\author[rug]{Jacquelien M.A. Scherpen}\ead{j.m.a.scherpen@rug.nl},
    \author[udeg]{D. del Puerto-Flores}\ead{dunstano.delpuerto@academicos.udg.mx},

	\address[rug]{University of Groningen}
    \address[udeg]{Universidad de Guadalajara}


	\begin{abstract}
	In this document, we deal with the stabilization problem of slow-fast systems (or singularly perturbed Ordinary Differential Equations) at a non-hyperbolic point. The class of systems studied here have the following properties: 1) they have one fast variable and an arbitrary number of slow variables, 2) they have a non-hyperbolic singularity of the fold type at the origin. The presence of the aforementioned singularity complicates the analysis and the controller design of such systems. In particular, the classical theory of singular perturbations cannot be used. We show a novel design process based on \emph{geometric desingularization}, which allows the stabilization of a fold point of singularly perturbed control systems. Our results are exemplified on an electric circuit.
\end{abstract}

\end{frontmatter}

\section{Introduction}

Slow-fast systems are characterized by having more than one timescale. There are many phenomena in nature that behave in two or more timescales such as population dynamics, cell division, electrical circuits, power networks, chemical reactions, neuronal activity, etc. \cite{KokotovicApps,KosiukS11,Galtier2012,Shilnikov2012,Kuehn2015}. A particular property of slow-fast systems (under certain hyperbolicity conditions) is that they have a structure suitable for model order reduction. Simply put, certain slow-fast systems can be decomposed into two simpler subsystems, the slow and the fast. The analysis of those two subsystems allows a complete understanding of the more complex and higher dimensional one. A mathematical theory supporting the previous fact is Geometric Singular Perturbation Theory (GSPT) \cite{Fenichel1979}, see also \cite{Kokotovic:1986:SPM:576779}. The above, however, relies on the strong assumption of global timescale separation. When that does not hold, then the classical technique of model order reduction cannot be employed. Thus, new mathematical tools need to be introduced in order to deal with problems without global timescale separation.

Regarding the latter situation, many interesting phenomena are characterized by not having a global timescale separation. This means that the variables of the system do not always have the same timescale relation throughout the phase-space. Mathematically speaking, this phenomenon is characterized by singularities of the critical manifold, see Section \ref{sec:preliminaries}, where the timescale separation does not hold; and in a qualitative sense, one usually observes jumps in the phase portrait of the slow-fast system. Such an effect is also called \emph{loss of normal hyperbolicity}. Prototypical examples of two-timescale systems without global timescale separation are the van der Pol oscillator \cite{vanderpol_heart}, neuronal models \cite{Shilnikov2012}, and electrical circuits with impasse points \cite{ChuaImpasse1,ChuaImpasse2,Reissig1}. Since loss of normal hyperbolicity is present in many models, there is an increased need of their accurate understanding. The distinction between hyperbolic\footnote{Global time scale separation.} and non-hyperbolic\footnote{No global timescale separation.} slow-fast systems is much more than qualitative. From an analysis point of view, the available techniques are quite different; while the hyperbolic case is well established, the non-hyperbolic scenario still presents many challenges. 

In the context of control systems, hyperbolic slow-fast systems and the related model order reduction are nowadays well understood and have been used in many applications, e.g. \cite{KokotovicApps,Spong1987,Sanfelice2011692,Saksena1984,Han20121904}. The main and powerful idea in the classical hyperbolic context is to design controllers for the reduced subsystems, which later guarantee stability of the overall slow-fast system. On the other hand, non-hyperbolic slow-fast system are far from being well understood, and to the best knowledge of the authors, there is not much progress yet. From a dynamical systems perspective, the technique called \emph{geometric desingularization} \cite{DumRou2,Krupa1} has been used to understand the complex behavior of slow-fast systems around non-hyperbolic points. In this regard, the authors have made preliminary progress in bringing such technique to the control systems community \cite{JardonMTNS2016,JardonACC2017} in the planar case. 

The main contribution of this document is the development of a control design method based on geometric desingularization. This provides a solution to the problem of the stabilization of non-hyperbolic singular perturbation problems. Although the technique presented here is completely different from the classical one \cite{Kokotovic:1986:SPM:576779}, the idea remains the same: to obtain simpler subsystems where the control design becomes more accessible. Furthermore, this document also generalizes our preliminary results of \cite{JardonMTNS2016,JardonACC2017} to systems with an arbitrary number of slow variables and one fast variable.

The rest of this document is organized as follows: in Section \ref{sec:preliminaries} we provide preliminary information regarding slow-fast systems followed by a description of the geometric desingularization method in Section \ref{sec:blowup}. In Section \ref{sec:setting} the specific problem under study is stated. Next, Section \ref{sec:geom}, we apply the geometric desingularization to control systems. Afterwards, in Section \ref{sec:design}, we develop a controller based on the method previously introduced. Interestingly, we show that it is possible to inject a hyperbolic behavior to a non-hyperbolic slow-fast control system even though the fast variable is not actuated. Later, in Section \ref{sec:examples} we exemplify our results for an electrical circuit. We finish in Section \ref{sec:conclusions} with some concluding remarks and a digression on open problems and future work.

\section{Preliminaries}\label{sec:preliminaries}

{\bfseries{Abbreviations:} } SFS stands for Slow-Fast System, SFCS for Slow-Fast Control System, and NH for Normally Hyperbolic.

{\bfseries{Notation:} } $\R$, $\mathbb Z$ and $\mathbb N$ denote the fields of real, integer, and natural numbers respectively. Given a field $\mathbb F$, $\mathbb F^n$ denotes the $n$-cross-product $\mathbb F\times \cdots\times \mathbb F$.  The symbols $\R_+$ and $\R_{\geq 0}$ are respectively used to denote the positive and the non-negative real numbers. $\I_n$ denotes the $n$-dimensional identity matrix. The dimension of the slow and fast variables is $n_s$ and $n_f$ respectively, and $N=n_s+n_f$. Given $x=(x_1,\ldots,x_n)\in\R^n$ and $\alpha=(\alpha_1,\ldots,\alpha_n)\in\mathbb Z^n$, we denote $x^\alpha=(x_1^{\alpha_1},\ldots,x_n^{\alpha_n})$.  The symbol $\s^n$ denotes the $n$-sphere. Given a matrix $A\in\R^{n\times m}$, $[A]_i$ denotes its $i$-th row. Whenever a matrix $A$ is square, $|A|$ denotes its determinant. The vector $\hat e_1$ denotes the canonical vector $\hat e_1=(1,0,\ldots,0)\in\R^n$. The parameter $\ve$ is always assumed $0<\ve\ll 1$. Let $\n$ be an $n$-dimensional manifold, a vector field $\X:\n\to T\n$ is written as $\X=\sum_{i=1}^n \X_i\parc{x_i}$, where $(x_1,\ldots,x_n)$ is a coordinate system in $\n$.

A slow-fast system (SFS) is a singularly perturbed ordinary differential equation of the form
\begin{equation}\label{sf1}
\begin{split}
\dot x &= f(x,z,\ve)\\
\ve\dot z &= g(x,z,\ve),
\end{split}
\end{equation}%
where $x\in\R^\ns$ (slow variable), $z\in\R^\nf$ (fast variable), $f$ and $g$ are sufficiently smooth functions, and the independent variable is the slow time $t$. One can also define a new time parameter $\tau=\frac{t}{\ve}$ called the fast time, and then \eqref{sf1} is rewritten as
\begin{equation}\label{sf2}
\begin{split}
x' &= \ve f(x,z,\ve)\\
z' &= g(x,z,\ve),
\end{split}
\end{equation}%
where the prime $'$ denotes derivative with respect to $\tau$. Note that \eqref{sf1} and \eqref{sf2} are equivalent as long as $\ve>0$. For convenience of notation we refer to \eqref{sf2} as $\X_\ve$, and write \eqref{sf2} the $\ve$-family of vector fields $\X_\ve=\ve f(x,z,\ve)\parcs{}{x}+g(x,z,\ve)\parcs{}{z}$.

A first step towards understanding the dynamics of \eqref{sf1} or \eqref{sf2} is to consider the limit equations when $\ve\to 0$. In such a limit \eqref{sf1} becomes a Differential Algebraic Equation (DAE) of the form
\begin{equation}\label{cde}
\begin{split}
\dot x &= f(x,z,0)\\
0&= g(x,z,0).
\end{split}
\end{equation}%
On the other hand $\X_\ve$  becomes
\begin{equation}\label{layer}
\X_0=0\parcs{}{x}+g(x,z,0)\parcs{}{z},
\end{equation}%
which is called \emph{the layer equation}. In principle, \eqref{cde} and \eqref{layer} are not related anymore, however, the critical manifold draws a bridge between them.

\begin{defn}The \emph{critical manifold} is defined as
\begin{equation}
\S=\left\{ (x,z)\in\R^\ns\times\R^\nf\,|\,g(x,z,0)=0 \right\}.
\end{equation}%
\end{defn}

Note that $\S$ is $\ns$-dimensional and serves as the phase-space of the DAE \eqref{cde} and as the set of equilibrium points of the layer equation $\X_0$. A very important property of critical manifolds is \emph{normal hyperbolicity}.

\begin{defn}\label{nh}
A point $s\in\S$ is called hyperbolic if it is a hyperbolic equilibrium point of $\X_0$. The manifold $\S$ is called normally hyperbolic (NH) if each point $s\in\S$ is a hyperbolic equilibrium point of $\X_0$.
\end{defn}
The importance of NH critical manifolds is explained by Geometric Singular Perturbation Theory (GSPT) \cite{Fenichel1979,Jones,Kaper,Kuehn2015}. Briefly speaking, the following hold.
\begin{itemize}[leftmargin=*]
	\item  Let $\S_0\subseteq\S$ be a compact, NH invariant manifold of a SFS. Then, for $\ve>0$ sufficiently small, there exists a locally invariant manifold $\S_\ve$ which is diffeomorphic to $\S_0$ and lies within distance of order $O(\ve)$ from $\S_0$.
	\item The flow of the SFS $X_\ve$ along $\S_\ve$ converges to the flow of the DAE along $\S_0$ as $\ve\to 0$.
	\item $\S_\ve$ has the same stability properties as $\S_0$.
	\item $\S_\ve$ is in general not unique. However, any two perturbations of $\S_0$ lie within exponentially small distance $O(\exp(-c/\ve))$, $c>0$, from each other. Any of the representatives of $\S_\ve$ is called \emph{the slow manifold}
\end{itemize}

On the other hand, we say that a SFS is non-hyperbolic if there exists a point $p\in\S$ such that the matrix $\parcs{g}{z}(p)$ has at least one eigenvalue with zero real part. The loss of normal hyperbolicity can be related to jumps or rapid transitions in, e.g., biological systems, climate models, chemical reactions, nonlinear electric circuits, or neuron models~\cite{KosiukS11,Krupa2,vanderpol_heart,Desroches1,Roberts1,Rotstein2013}. 

The contribution of this paper is the stabilization of a non-hyperbolic point of a control system with two timescales. Before presenting the main tool to be used, let us recall the classical (hyperbolic) setting of control of singularly perturbed systems.

\subsection{Composite control of slow-fast control systems}\label{sec:SFCS}
Let us define a slow-fast control systems (SFCS's) as
	\begin{equation}\label{def:cSFS}
		\begin{split}
			\dot x &= f(x,z,\ve,u)\\
			\ve\dot z &= g(x,z,\ve,u),
		\end{split}
	\end{equation}%
where $x\in\R^\ns$, $z\in\R^\nf$, and $u\in\R^m$ is a control input. Let us now briefly recall a classical method to design controllers for \emph{hyperbolic SFCS's} \cite{Kokotovic:1986:SPM:576779}. Assume that the critical manifold of \eqref{def:cSFS} is normally hyperbolic in a compact set $(x,z,u)\in U_x\times U_z\times U_u\subset \R^\ns\times\R^\nf\times\R^m$. The goal is to design a control $u$ that stabilizes the origin $(x,z)=(0,0)\in\R^\ns\times\R^\nf$ for $\ve>0$ sufficiently small.  The hyperbolicity assumption implies that the critical manifold $\S=\left\{ g(x,z,u)=0\right\}$ can be locally expressed as a graph $z=h(x,u)$ for $(x,u)\in U_x\times U_u$. The idea of composite control is to design $u$ as a sum of two simpler controllers, namely $u=u_s+u_f$ where $u_s=u_s(x)$ is ``the slow controller'' and $u_f=u_f(x,z)$ is ``the fast controller''. When designing $u$, $u_f$ must be chosen so that it does not destroy the normal hyperbolicity of the system, meaning that $g(x,z,u_s(x)+u_f(x,z))=0$ must have (locally) a unique root $z=H(x)$. Thus, it is often required that the effects of the fast controller $u_f$ disappear along the $\S$, that is $u_f|_{\S}=0$. In this way $z=h(x,u_s(x))$ is (locally) a unique root of $g(x,z,u)=0$ and the reduced flow along $\S$ is given by 
\begin{equation}\label{h-red-1}
	\dot x =f(x,H(x),u_s(x)).
\end{equation}
Note that the reduced system is, as expected, independent of the fast variable $z$ and of the fast controller $u_f$. Thus, the slow controller $u_s$ is designed to make $\left\{x=0\right\}$ an asymptotically equilibrium point of \eqref{h-red-1}. After $u_s$ has been designed, one studies the layer problem $z'=g(x,z,u_s(x)+u_f(x,z))$, where $x$ is taken as a fixed parameter, and where the fast controller is designed so that $z=h(x,u_s(x))$ is a set of asymptotically stable equilibrium points. The previous strategy plus some extra (technical) interconnection conditions guarantee that the origin $(x,z)=(0,0)\in\R^{n_s}\times\R^{n_f}$  is an asymptotically stable equilibrium point of the closed-loop system \eqref{def:cSFS} for $\ve>0$ but sufficiently small \cite{Kokotovic:1986:SPM:576779,Kokotovic1976,Isidori1,Marino1988}. This feature of normal hyperbolicity has been exploited in many applications, a few examples are \cite{Marszalek2,Tang1,Sanfelice2011692,Pan2015,Zhou1,Vecchio1,Gajic20011859,Nesic1}.

\section{The Geometric Desingularization method}\label{sec:blowup}

It is evident that the composite control strategy described in Section \ref{sec:SFCS} is not applicable around non-hyperbolic points. 
There have been already some efforts to study the stabilization problem of non-hyperbolic points of nonlinear systems, see \cite{Marconi1,Menini2010537}. However, these do not address nonlinear two timescales systems, where many open problems remain to be solved. We propose to use the \emph{geometric desingularization} method to design controllers in a rather simple and standard way for multiple timescale systems around non-hyperbolic points. 

Before going into details, note that $\X_\ve$ \eqref{sf2} is an $\ve$-parameter family of $N$-dimensional\footnote{Recall that in this document $N=n_s+n_f$.} vector fields. For their analysis it is more convenient to lift such family up and consider instead a single ($N+1$)-dimensional vector field defined as
\begin{equation}\label{eq:X}
\X:\begin{cases}
	x' &= \ve f(x,z,\ve)\\
	z' &= g(x,z,\ve)\\
	\ve'&=0
\end{cases}	
\end{equation}
The \emph{geometric desingularization} method, also known as blow up, is a geometric tool introduced in \cite{DumRou2} for the analysis of SFSs around non-hyperbolic points, see also \cite{JardonThesis,Jardon3,KosiukS11,Krupa2,Krupa20102841,Szmolyan2004RelaxationR3,Szmolyan2001419,Kuehn2015}. In an intuitive way, the blow up method transforms non-hyperbolic points of SFSs to (partially) hyperbolic ones. Below we provide a brief introduction to the technique, further details can be found in \cite{DumRou2,Kuehn2015} . 

\newcommand{\bz}{\bar{z}}
\newcommand{\bx}{\bar{x}}
\newcommand{\br}{\bar{r}}
\newcommand{\be}{\bar{\ve}}
\renewcommand{\B}{\mathcal{B}}
\newcommand{\Z}{\mathcal{Z}}

\begin{defn}  A (quasi-homogeneous)\footnote{A homogeneous blow up (or simply blow up) refers to all the exponents $\alpha$, $\beta$, $\gamma$ set to $1$.  } blow up transformation is a map
\begin{eqnarray}
	\Phi &:& \s^{N}\times \R\to\R^{N+1}\nonumber\\
	&&\Phi(\bx,\bz,\be,\br)\mapsto(\br^{\alpha}\bx,\br^{\beta}\bz,\br^\gamma\be),
\end{eqnarray}%
where  $(\bx,\bz,\be)\in \s^N$, $\alpha\in\mathbb Z^\ns$, $\beta\in\mathbb Z^\nf$ and $\gamma\in\mathbb Z$.goo
\end{defn}

In many applications, and in particular in this paper, it is enough to consider $\br\in\R_{\geq 0}$, which implies $\be\in\R_+$. Thus, let $\B=\s^{N}\times \R_{\geq 0}$, and $\Z=\s^{N}\times \left\{ 0 \right\}$. A blow up induces a vector field on $\B$ as follows.

\newcommand{\bX}{\bar{\mathcal{X}}}

\begin{defn} Let $\X:\R^{N+1}\to T\R^{N+1}\simeq\R^{N+1}$ be a smooth vector field, and $\Phi:\B\to\R^{N+1}$ a blow up map. The \emph{blow up} of $\X$ is a vector field $\bX:\B\to T\B$ induced by $\Phi$ in the following sense
\begin{equation}
	\bX=D\Phi^{-1}\circ \X \circ \Phi.
\end{equation}
\end{defn}
\newcommand{\tX}{\tilde{\mathcal{X}}}
It may happen that the vector field $\bX$ degenerates along $\Z$. In such a case one defines the \emph{desingularized vector field} $\tX$ as
\begin{equation}
	\tX=\frac{1}{\bar r^m}\bX,
\end{equation}
for some well suited $m\in\N$ so that $\tX$ is not degenerate, and is well defined along $\Z$. Note that the vector fields $\bX$ and $\tX$ are equivalent on $\s^N\times \left\{ \bar r>0\right\}$. Moreover, if the weights $(\alpha,\beta,\gamma)$ are well chosen, the singularities of $\tX|_{\bar r=0}$ are partially hyperbolic or even hyperbolic, making the analysis of $\tX$ simpler than that of $\X$. Due to the equivalence between $\X$ and $\tX$, one obtains all the local information of $\X$ around $0\in\R^{N+1}$ from the analysis of $\tX$ around $\Z$.

While doing computations, it is more convenient to study the vector field $\tX$ in charts. A chart is a parametrization of a hemisphere of $\B$.

\renewcommand{\K}{\kappa}

\begin{defn} A chart is obtained by setting one of the coordinates $(\bx,\bz,\be)\in\s^N$ to $\pm 1$ in the definition of $\Phi$. If such coordinate is, for example, $\bx_i$, then the chart is denoted as $\K_{\pm \bx_i}$.
\end{defn}

For example, the chart $\K_{\be}$ is defined by the coordinates $(\br^\alpha\bx,\br^\beta\bz,\br^\gamma)$, while the chart $\K_{-\bx_1}$ by $(-\br^{\alpha_1},\br^{\alpha_2}\bx_2,\ldots,\br^{\alpha_n}\bx_n,\br^\beta\bz,\br^\gamma\be)$. Similarly to the blow up map, a chart $\K_{\pm w}$ induces a vector field to be denoted by $\tX^{\pm w}$. Note that each chart covers only part of $\B$ but all of them define an open cover of $\B$. Therefore, one uses several charts in order to have a full understanding of the dynamics of $\tX$ around $\B$. These charts, together with their vector fields, have to be ``patched'' together via transition maps, defined as follows.

\begin{defn}\label{def:trans} Let $\K_v$ and $\K_w$ be charts such that $\K_v\cap \K_w\neq\emptyset$. A \emph{transition map} $\phi_{vw}$ is a strictly positive smooth function, well defined in $\K_v\cap \K_w$, which makes the following diagram commute.
\begin{center}
	\begin{tikzpicture}
  \matrix (m) [matrix of math nodes,row sep=3em,column sep=4em,minimum width=2em]
  {
      & \B & \\
      \R^{N+1} & & \R^{N+1} \\};
  \path[-stealth]
    (m-1-2) edge node[above, midway] {$\K_v$} (m-2-1)
	(m-1-2) edge node[above, midway] {$\K_w$} (m-2-3)
	(m-2-1) edge node[above, midway] {$\phi_{vw}$} (m-2-3)
	;
\end{tikzpicture}
\end{center}
	
\end{defn}

\begin{figure}[b!]\centering
	\begin{tikzpicture}[scale=0.75]
		\pgftext{\includegraphics{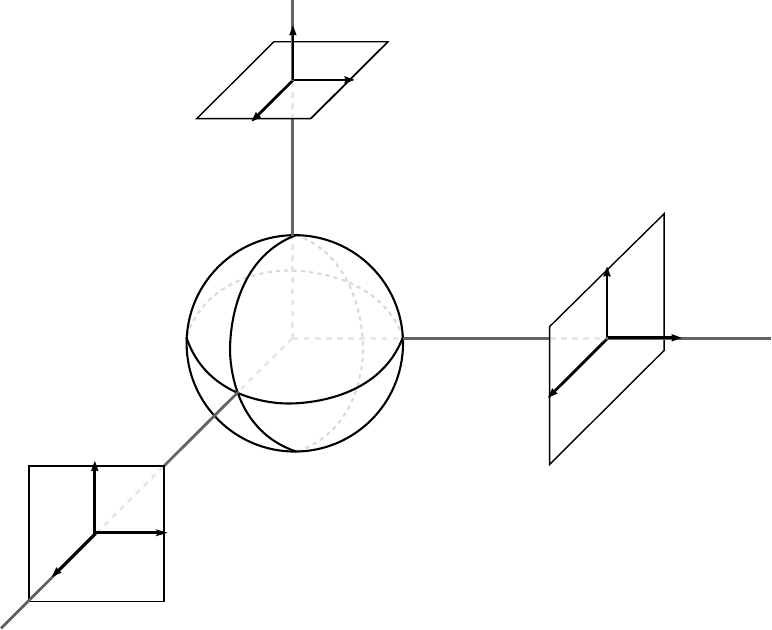}}
		\node at (-2,-1) {\small$\bar \ve$};
		\node at (0.3,0) {\small$\bar x$};
		\node at (-0.7,1) {\small$\bar z$};

		\node at (-3,-1.35) {\small $z_1$};
		\node at (-2.,-2.25) {\small $x_1$};
		\node at (-3.15,-2.75) {\small $r_1$};

		\node at (2.3,.85) {\small $z_2$};
		\node at (1.45,-.9) {\small $\ve_2$};
		\node at (3.15,-.1) {\small $r_2$};

		\node at (-.1,2.3) {\small $x_3$};
		\node at (-1.45,1.75) {\small $\ve_3$};
		\node at (-.7,3.) {\small $r_3$};

	\end{tikzpicture}
	\caption{Illustration of a blow up and three charts with corresponding local coordinates.}
	\label{f:bu}
\end{figure}
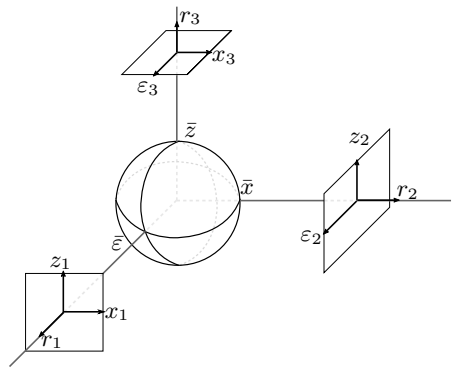

A schematic of the above definitions is shown in Figure \ref{f:bu}. 

In practice, we do not need to study all the charts covering $\B$ but rather a few of them. A good way to choose the relevant charts is to first perform a qualitative analysis of the dynamics of a SFS to realize which charts provide useful local information. After one has studied the vector fields in the charts and has connected the information in ``adjacent'' charts, one performs a blow down, the inverse of the blow up. In summary, the steps to follow when studying the dynamics of a SFS near a non-hyperbolic point and via the geometric desingularization method are as follows.

\begin{enumerate}[leftmargin=*]
	\item  Find a suitable blow up. This is, find the proper weights of the blow up map and rescaling so that the non-hyperbolic point is desingularized.
	\item  Find which charts are relevant and express the blow up in the respective local coordinates.
	\item  Investigate the dynamics of the local vector fields in the charts.
	\item  Connect the local results via transition maps and blow down.
\end{enumerate}

\section{Setting of the problem}\label{sec:setting}

The forthcoming sections are dedicated to the stabilization of a particular class of SFCS. We consider systems with one fast variable ($n_f=1$) and an arbitrary number of slow variables ($n_s$). Preliminary results for the case $(n_s,n_f)=(1,1)$ are presented in \cite{JardonMTNS2016,JardonACC2017}. Here we extend such results by allowing an arbitrary number of slow variables. To start, we consider that the critical manifold has a particular geometric structure.

\begin{defn}[Fold point] Consider the SFS \eqref{eq:X} and let $p=(x_0,z_0)\in\S$. If 
\begin{equation}
\begin{split}
g(x_0,z_0,0) &=0,\quad \frac{\partial g}{\partial z}(x_0,z_0,0)=0,\\
\frac{\partial^2 g}{\partial z^2}(x_0,z_0,0) &\neq 0,
\end{split}
\end{equation}
then $p$ is called a fold point of the SFS.
\end{defn}

\newcommand{\M}{\mathcal{M}}


We shall study a SFCS whose critical manifold (in open loop) has a fold point at the origin. We do this by defining the system
\begin{equation}\label{eq:c-fold-1}
	\begin{split}
		\dot x &= f(x,z,\ve,u)\\
		\ve\dot z &= g(x,z,\ve,u),
	\end{split}
\end{equation}%
where $x=(x_1,\ldots,x_\ns)\in\R^\ns$, $z\in\R$, $u\in\R^m$, and $g$ satisfies\footnote{In a more singularity theory language, $g=0$ coincides with the critical set of the Fold catastrophe \cite{Arnold_singularities,jardon2014polynomial,takens1976constrained}. Moreover, the choice of the negative sign is just for convenience and a completely equivalent analysis follows otherwise.} $g(x,z,0,0)=-(z^2+x_1)$. The choice of \eqref{eq:c-fold-1} is due to: 1) Fold singularities are generic in one-parameter families of smooth functions \cite{Arnold_singularities,Brocker}, therefore, they are expected to appear in nonlinear SFSs with at least one slow variable. 2) In many applications, jumps and rapid transitions occur through fold points, the best-known example are relaxation oscillations in the van der Pol oscillator \cite{vanderpol_heart}, although other examples can be found in e.g. quantum electronics \cite{neumann2003slow}, neuroscience \cite{Desroches1,Shilnikov2012} or biochemistry \cite{Szmolyan2004RelaxationR3}.


 We shall design a feedback controller $u=u(x,z,\ve)$ that stabilizes the origin of \eqref{eq:c-fold-1}. Moreover, we consider the \emph{slow-actuated} (see Remark \ref{rem:slow} below)  and affine case, namely
\begin{equation}\label{eq:c-fold-2}
	\begin{split}
		\dot x &= f(x,z,\ve)+B(x,z,\ve) u(x,z,\ve)\\
		\ve\dot z &= g(x,z,\ve)=-(z^2+x_1),
	\end{split}
\end{equation}%
where $B=B(x,z,\ve)\in\R^{\ns\times m}$ depends smoothly on its arguments. We make the following two assumptions.

\begin{enumerate}[leftmargin=*]
	\item[{\bfseries{A1.}}] The matrix $B$ is invertible. Therefore $m=n_s$. Such an assumption implies that the slow-dynamics are fully actuated.
	\item[{\bfseries{A2.}}] $f(x,z,\ve)=f_0(x,z)+\ve f_1(x,z,\ve)$ with $f_1(0,0,\ve)=0$. This ensures that $\ve$ does not introduce any constant drift near the origin.
\end{enumerate}


\begin{rem} \label{rem:slow}
The fully actuated case is in fact solvable by a minor modification of the composite control method: 1) propose the controller $u=U_h(z)+U_c(x,z)$, where the purpose of $U_h$ is to make the closed-loop critical manifold NH. 2) Once the critical manifold is NH, one proceeds by designing $U_c$ as in the composite control (recall Section \ref{sec:SFCS}).
\end{rem}


\section{Geometric desingularization of a folded slow-fast control system}\label{sec:geom}

	In this section we perform the geometric desingularization of the system
\begin{equation}\label{eq:g-fold-0}\X:
	\begin{cases}
		x' &= \ve \left( f(x,z,\ve)+B(x,z,\ve) u \right)\\
		z' &= -(z^2+x_1)\\
		\ve'&=0,
	\end{cases}
\end{equation}%
which is equivalent to \eqref{eq:c-fold-2}. Notice that $\X$ is a vector field on $\R^{n_s+2}$. Without loss of generality we can write $f= A+L(x,z)+F(x,z,\ve)$,
where $A=f(0,0,0)$, $L(x,z)$ is a linear map, i.e., $L(x,z)=L_1x+L_2z$ with $L_1\in\R^{\ns\times \ns}$, $L_2\in\R^\ns$,  and $F(x,z,\ve)$ stands for all the higher order terms and satisfies, due to Assumption {\bfseries{A2}}, $F(0,0,\ve)=0$. Thus, \eqref{eq:g-fold-0} is rewritten as
\begin{equation}\label{eq:g-fold-1}\X:
	\begin{cases}
		x' &= \ve \left( A+L(x,z)+B(x,z,\ve) u+F(x,z,\ve) \right)\\
		z' &= -(z^2+x_1)\\
		\ve'&=0.
	\end{cases}
\end{equation}%
The blow up map $\Phi:\s^{\ns+1}\times [0,\infty)\to\R^{\ns+2}$ is defined by
\begin{equation}\label{bbuu}
	x=\br^\alpha \bx, \; z=\br^\gamma\bz, \; \ve=\br^\rho\be,
\end{equation}%
where $\bx=(\bx_1,\ldots,\bx_\ns)$, $\sum_{i=1}^\ns \bx_i+\bz+\be=1$,  $\alpha\in\mathbb Z^\ns$, $(\gamma,\rho)\in\mathbb Z^2$ and $\br\in\R_{\geq0}$. In principle, we could set all the exponents in the blow up map to $1$, but this would require more than one coordinate transformation to completely desingularize \eqref{eq:g-fold-1}  \cite{structures,JardonThesis,Kuehn2015}.  A good choice of weights is\footnote{The choice of the weights depends on the quasihomogenity type of the vector field, for details see \cite{ArnoldDS6,Sto10,JardonThesis,jardon2015formal}.} $\alpha_1=\cdots=\alpha_n=2$, $\gamma=1$, and $\rho=3$. As described in Section \ref{sec:blowup}, it is more convenient to work with charts, which in this case are defined as $\K_{\pm\bx_i}=\left\{ \bx_i=\pm 1\right\}$, $\K_{\pm\bz}=\left\{ \bz=\pm 1\right\}$, and $\K_{\be}=\left\{ \be= 1\right\}$, and its induced local vector fields. To simplify the exposition, let us adopt the following notation.

\newcommand{\tX}{\tilde{\mathcal{X}}}
\newcommand{\Bx}{\mathbf{X}}
\newcommand{\Bxi}{\mathbf{X}^{\pm i}}
\newcommand{\bB}{\bar{B}}
\newcommand{\bF}{\bar{F}}

{\bfseries{Notation:}} Let $\K_{\pm w}$ be a chart. 
If $H(x,z,\ve)$ is a function in the original space ($(x,z,\ve)\in\R^{\ns+2}$) its blow up is denoted by $\bar H$, i.e., $\bar H=H\circ\K_{\pm w}$. 
\newcommand{\bxi}{\bar{x}}

\begin{rem}
	In the analysis to be performed below, the most important chart is $\K_{\be}$. This is because it is precisely in such a chart where the singular dependence on $\ve$ is resolved. 
    Next we provide the blown up vector field in the chart $\K_{\be}$.
\end{rem}

\begin{prop}\label{prop:blown-vfs} The vector field $\tX^{\be}$, which corresponds to the blow up of \eqref{eq:g-fold-1} in the chart $\K_{\be}$, reads as
%
%
%
%
%
%
%
%
\begin{equation}\label{eq:Xe}
	\tX^{\be}:\begin{cases}
		\br' &=0\\
		\bx' &=  A + \br^2L_1 \bx +\br L_2 \bz +\bar B\bu +\bar F \\
		\bz' &=-(\bz^2+\bx_1),
	\end{cases}   
\end{equation}
where $\bu=\bu(\br,\bx,\bz)$ and $\bF=\bF(\br,\bx\,\bz)$. 
\end{prop}
\begin{pf} The coordinates in the chart $\K_{\be}$ are defined by $(x_1,\ldots,x_\ns,z,\ve)=(\br^2\bx_1,\ldots,\br^2\bx_\ns,\br\bz,\br^3)$. From $\ve=\br^3$, it follows that $\br'=0$.
%
Then, $x' = 2\br\br'\bx + \br^2\bx' $, which implies
\begin{equation}\label{ex}
	\bx'= \br\left( A + \br^2L_1 \bx  +\br L_2 \bz +\bB\bu +\bar F \right),
\end{equation}
where $\bar B=\bar B(\br,\bx,\bz)=B(\br^2\bx,\br\bz,\br^3)$, and similarly for $\bu$ and $\bF$. In a similar way one obtains
\begin{equation}\label{ez}
	\bz'=-\br\left( \bz^2+\bx_1 \right).
\end{equation}
Note, however, that the blown up vector fields \eqref{ex}-\eqref{ez} degenerate at $\left\{ \br=0 \right\}$. Thus, one defines the desingularized vector field by rescaling by a factor $\frac{1}{\br}$, which leads to the result.
%
\hfill\ensuremath\qed
\end{pf}

\begin{rem} The local vector fields in the other charts are obtained similarly to Proposition \ref{prop:blown-vfs}.
\end{rem}

The stability properties of the blown up vector field $\tX^{\be}$ of \eqref{eq:Xe} are carried over similar properties into the original SFS. The main argument is that ``stability should be invariant under changes of coordinates'', as is analyzed in the next section.

\section{Controller design via geometric desingularization}\label{sec:design}

\newcommand{\br}{\bar{r}}
\newcommand{\bx}{\bar{x}}
\newcommand{\by}{\bar{y}}
\newcommand{\bz}{\bar{z}}
\newcommand{\be}{\bar{\ve}}
\renewcommand{\K}{\kappa}
\newcommand{\X}{\mathcal{X}}
In this section we show how to use geometric desingularization to stabilize a non-hyperbolic point of a SFCS. The method is to design a controller in the blown up space and then to blow such controller down. We do this mainly in the central chart  $\K_{\be}$ as argued above. However, the directional charts may be used for other purposes, for example, to increase performance or to shape the transitory behavior of the trajectories of a SFS, etc.

\renewcommand{\br}{ \rho}
\renewcommand{\bx}{ \chi}
\renewcommand{\bz}{ \zeta}
\newcommand{\bu}{\bar u}
\newcommand{\bF}{\bar F}
\newcommand{\bxi}[1]{\bx_{#1}}
\newcommand{\tX}{\tilde{\mathcal X}}

To avoid confusion of notation, the local coordinates in $\K_{\be}$ are denoted as $(\bx,\bz,\br)=(\bxi{1},\ldots,\bxi{n},\bz,\br)\in\R^{\ns+2}$. Therefore the blow up map reads as
\begin{equation}
	(x_1,\ldots,x_\ns,z,\ve)=(\br^2\bx_1,\ldots,\br^2\bx_\ns,\br\bz,\br^3),
\end{equation}
and then, from Proposition \ref{prop:blown-vfs} $\tX^{\be}$ reads as
\begin{equation}\label{eq:C-main}
	\tX^{\be}:\begin{cases}
		\br' &=0\\
		\bx' &=  A + \br^2L_1\bx +\br L_2\bz+\bar B \bu +\bF \\
		\bz' &=-(\bz^2+\bxi{1}),
	\end{cases}
\end{equation}
where $\bar B=\bar B(\br,\bx,\bz)=B(\br^2\bx,\br\bz,\br^3)$, and similarly for $\bF$ and $\bu$. Moreover $\bF|_{\left\{ \br=0\right\}}=0$. Observe that Assumption {\bfseries{A1}} readily implies local controllability of $\bx' =  A + \br^2L_1\bx +\br L_2\bz+\bar B \bu +\bF$ for all $\rho\geq 0$ sufficiently small.
\begin{rem}
	Note that in \eqref{eq:C-main} $\br$ is a regular perturbation parameter. Thus, the idea is to solve first the stabilization problem for $\br=0$ and then use regular perturbation arguments \cite{murdock1999perturbations} to guarantee the stability of the origin of \eqref{eq:C-main} for $\br\geq 0$ sufficiently small.
\end{rem}

The main argument to relate the stability of the blown up vector field $\tX^{\be}$ and $\X$ is the following.


\begin{prop}\label{cor:main2} Consider a SFS $\X$ given by \eqref{eq:g-fold-1}
and its blow up version $\tX^{\be}$ in the chart $\K_{\be}$ given by \eqref{eq:Xe}.
If for each $\br\in(0,\br_0]$ the point $(\bx,\bz)=(0,0)$ is G.A.S. for $\tX^{\be}$, then for each $\ve\in(0,\br^{1/3}]$ the point $(x,z)=(0,0)$ is  G.A.S. for $\X$.
\end{prop}
\begin{pf} Let $\br=\br_0>0$ be fixed. Then the change of coordinates is defined as
\begin{equation}\label{eq:phi}
\begin{split}
\phi(\bx_1,\ldots,\bx_\ns,\bz) &=(\br_0^2\bx_1,\ldots,\br_0^2\bx_\ns,\br_0\bz)\\
&=(x_1,\ldots,x_\ns,z).
\end{split}
\end{equation}
Note that $\phi$ is a diffeomorphism with positive definite Jacobian. On the other hand, the hypothesis that $(\bx,\bz)=(0,0)$ is G.A.S. for $\tX^{\be}$ implies that there exists a $\br$-family of Lyapunov functions $\bar V_{\br}(\bx,\bz)$ satisfying
 \begin{itemize}[leftmargin=*]
 \item $\bar V_{\br}>0$ for all $(\bx,\bz)\in\R^{\ns+1}\backslash \left\{ 0\right\}$,
 \item $\bar V'_{\br}<0$ for all $(\bx,\bz)\in\R^{\ns+1}\backslash \left\{ 0\right\}$,
 \item $\bar V_{\br}$ is radially unbounded.
 \end{itemize}
Define the $\ve$-family of Lyapunov candidate functions $V_{\ve}=\bar V_{\rho}\circ\phi^{-1}$, where $\ve=\br^3$. From the properties of $\phi$ \eqref{eq:phi}, namely that $\phi$ is a diffeomorphism with positive definite Jacobian, it follows that $V_{\ve}>0$ and $V'_{\ve}<0$ for all $(x,z)\in\R^{\ns+1}\backslash \left\{ 0\right\}$. Therefore, $V_{\ve}$ is an $\ve$-family of Lyapunov functions. Finally, let $||(\bx,\bz)||\to\infty$, which clearly implies that $||(x,z)||=|| \phi(\bx,\bz) ||\to\infty$. From the definition of $V_\ve$ we have $V_\ve(x,z)=\bar V_\br\circ\phi^{-1}(x,z)=\bar V_\br(\ve^{-2/3}x,\ve^{-1/3}z)$, which in fact shows that $V_{\ve}$ is also radially unbounded. \ensuremath\hfill\qed
\end{pf}
Proposition \ref{cor:main2} means that we can design controllers to  stabilize a SFCS by designing it in the blown up space. The way the controller $\bu$ is actually designed depends on the specific context of the problem. Below we present a particularly interesting case where even though the origin is non-hyperbolic and the fast variable is not actuated, one is able to inject a hyperbolic behavior through actuating the slow variables.

\subsection{Hyperbolicity injection}

In this section we design a controller which induces a hyperbolic behavior in \eqref{eq:C-main} around the origin. We do this via the backstepping algorithm \cite{SastryBook}.


\newcommand{\bB}{\bar B}
\newcommand{\bBz}{\bar B_0}

\begin{prop}\label{p:ada} Consider \eqref{eq:C-main}. If $\bu$ is designed such that
\begin{equation}\label{pu1}
		\bB_0\bu=-A-c_1\bx+(c_0c_1+\gamma)\bz\hat e_1+\beta(\bx,\bz)\hat e_1,
\end{equation}
where $\bBz=\bB|_{\left\{\br=0\right\}}$, and
\begin{equation}\label{pu2}
	\bar\beta(\bx,\bz)=(2\bz-c_0)(\bz^2+\bxi{1})-c_1\bz^2,
\end{equation}
with $0<\gamma\ll c_1$ and $0<c_0\ll c_1$, then the origin of \eqref{eq:C-main} is rendered globally asymptotically stable for $\br\geq 0$ sufficiently small.

\end{prop}	

\newcommand{\buz}{\bu_0}

\begin{pf} First, we restrict the analysis to $\left\{ \br=0 \right\}$. Therefore we have that \eqref{eq:C-main} reads as
\begin{equation}\label{adp-1}
	\tX^{\be}|_{\br=0}=\left(A +\bBz \buz\right)\parc{\bx}-(\bz^2+\bxi{1})\parc{\bz},
\end{equation}
where $\buz$ denotes the restriction $\bu|_{\br=0}$.  Next, consider the system
\begin{equation}\label{ad-au3}
	\bz' =-(\bz^2+\bxi{1}).
\end{equation}
The idea is to first treat $\bxi{1}$ in \eqref{ad-au3} as a virtual controller which injects a hyperbolic behavior on $\bz$ in \eqref{ad-au3}. For this let $\bxi{1}=\alpha(\bz)$ and consider the Lyapunov function $V_0(\bz)=\frac{1}{2}\bz^2$. It follows that $V_0'=\bz\bz'=-\bz(\bz^2+\alpha)$. Define $\alpha(\bz)=-\bz^2+c_0\bz$ with $c_0>0$. Then, by Lyapunov arguments it follows that $\bz=0$ is a global asymptotic equilibrium point of the (virtual) closed-loop system $\bz' =-(\bz^2+\alpha)=-c_0\bz$. Note that, indeed, $\bz=0$ is a \emph{hyperbolic equilibrium point} of $\bz' =-(\bz^2+\alpha)$.

To simplify notation let $\xi=\bxi{1}-\alpha$, $Y=(Y_1,\ldots,Y_\ns)=(\xi,\bxi{2},\ldots,\bxi{\ns})\in\R^\ns$ and $W=(2\bz-c_0)(\xi+c_0\bz)\hat e_1$. Then \eqref{adp-1} is rewritten as
\begin{equation}\label{adp-2}
	\begin{split}
		Y' &=A-W+\bBz\buz\\
		\bz'&=-(c_0\bz+Y_1).
	\end{split}
\end{equation}
Let us now propose the Lyapunov function $V_1(Y,\bz)=\frac{1}{2}\gamma\bz^2+\frac{1}{2}Y^TY$, $\gamma>0$, it follows that
\begin{equation}
	V_1'=-c_0\bz^2+Y^T\left( A-W-\gamma\bz\hat e_1+\bB_0\buz \right).
\end{equation}
Thus we design $\buz$ so that 
\begin{equation}\label{eqproofu}
\bB_0\buz=-c_1 Y -A+W+\gamma\bz \hat e_1
\end{equation}
with $c_1>0$, and then $V_1'=-c_0\bz^2-c_1Y^TY$. Therefore, such choice of controller ensures that the origin $(Y,\bz)=(0,0)\in\R^{n+1}$ is a GAS equilibrium point of \eqref{adp-2}. Moreover, note that $(\xi,\bz)\to(0,0)$ imply $\bxi{1}\to 0$, which shows the claim. Finally, for the above algorithm to work, it is convenient to ensure that $\xi=\bxi{1}-\alpha \to 0$ fast enough. This is because it is precisely at $\xi=0$ when the dynamics of $\bz$ become hyperbolic. For the latter to happen, let us consider the closed-loop matrix associated to \eqref{adp-2}\footnote{In fact, the closed-loop system \eqref{adp-2} is linear.}, which reads as  
\begin{equation}
	\Phi=\begin{bmatrix}
		-c_1\I_n & \gamma\hat e_1\\
		-\hat e_1^T & -c_0
	\end{bmatrix}\in\R^{(\ns+1)\times(\ns+1)}.
\end{equation}
It is straightforward to show that the eigenvalues of $\Phi$ are
\begin{equation}
\left\{ \underbrace{-c_1,\ldots,-c_1}_{n_s-1}, \lambda_{1,2}\right\},
\end{equation}
where $\lambda_{1,2}\neq -c_1$ are the solutions of the quadratic polynomial $s^2+(c_0+c_1)s+\gamma+c_0c_1=0$. Both eigenvalues $\lambda_{1,2}$ have negative real part, so the closed-loop system \eqref{adp-2} is GAS, however, as argued above, it is convenient that $\xi=\bxi{1}-\alpha\to 0$ fast enough so that the backstepping algorithm quickly induces the desired hyperbolic behavior in $\bz$. To ensure this, note that if $\gamma$ is small, the eigenvalues of $\Phi$ are approximately $\left\{ \underbrace{-c_1,\ldots,-c_1}_{\ns\,\text{times}},-c_0 \right\}$, which motivates the condition $0<c_0\ll c_1$.  The expressions \eqref{pu1}-\eqref{pu2} are obtained by returning from \eqref{eqproofu} to the coordinates $(\bx,\bz)$. Since \eqref{eq:C-main} is a regular perturbation problem in $\br$, and the origin is still an isolated equilibrium point of \eqref{eq:C-main} for $\br>0$, all the above arguments hold for $\br>0$ sufficiently small \cite{murdock1999perturbations}.\hfill\ensuremath\qed
\end{pf}
The final step is to obtain the controller induced by Proposition \ref{p:ada}.
\begin{thm}\label{thm:mainAda}Consider the SFCS
\begin{equation}\label{eq:thm2}\X:
	\begin{cases}
		x' &= \ve \left( f(x,z,\ve)+Bu(x,z,\ve) \right)\\
		z' &= -(z^2+x_1)\\
		\ve'&=0,
	\end{cases} 
\end{equation}
If the control law $u$ is designed such that 
\begin{equation}
	 B_0 u =-A-\frac{c_1}{\ve^{2/3}}x+\frac{(c_0c_1+\gamma)}{\ve^{1/3}}z\hat e_1+\beta(x,z)\hat e_1
\end{equation}
where $A=f(0,0,0)$, $B_0=B(0,0,0)$ and
\begin{equation}
	\beta=\ve^{-2/3}\left( -c_1 z^2+(2\ve^{-1/3}z-c_0)(z^2+x_1) \right),
\end{equation}
with $0<c_0\ll c_1$ and $\gamma>0$, then the origin $(x,z)=(0,0)\in\R^{n+1}$ is rendered globally asymptotically stable for $\ve>0$ sufficiently small.
\end{thm}
\begin{pf}  
	The proof follows from Proposition \ref{p:ada}, the corresponding blow down of the controller $\bu$, and Proposition \ref{cor:main2}. \hfill\ensuremath\qed
\end{pf}

\begin{rem} In Theorem \ref{thm:mainAda} we only require the knowledge of $A=f(0,0,0)$, while the values of $L$ and $F$ are not important as long as they are bounded. Theorem \ref{thm:mainAda} can be extended to an adaptive, but local, version, where the value of $A$ may be assumed unknown, see \cite{JardonACC2017}. 
\end{rem}

\label{sec:ada}


\section{Example: Stabilization at a fold point of a nonlinear electric circuit}\label{sec:examples}

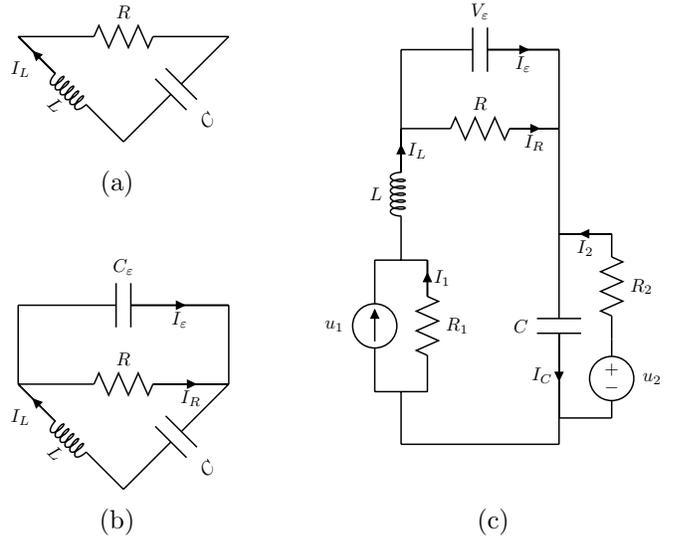
\begin{figure}[b!]\centering
\begin{tikzpicture}

\draw[white] (-2,2.5) rectangle (7,-5.5);

\node at (-0.5,1) {\pgftext{\includegraphics[scale=0.7]{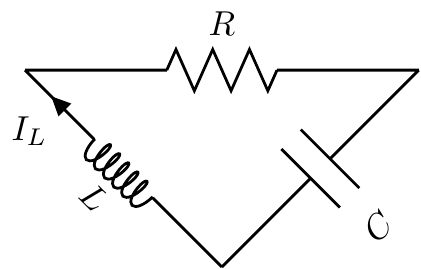}}};
\node at (-0.5,-0.5) {(a)};

\node at (-0.5,-3) {\pgftext{\includegraphics[scale=0.7]{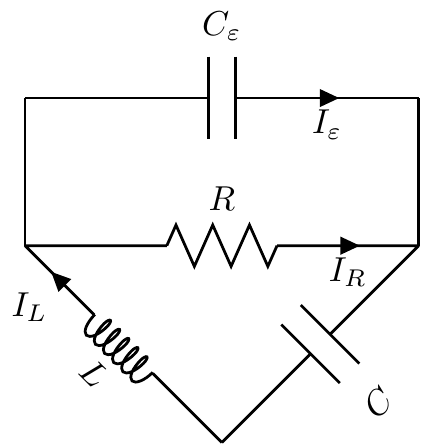}}};
\node at (-0.5,-5) {(b)};

\node at (4.5,-1) {\pgftext{\includegraphics[scale=0.7]{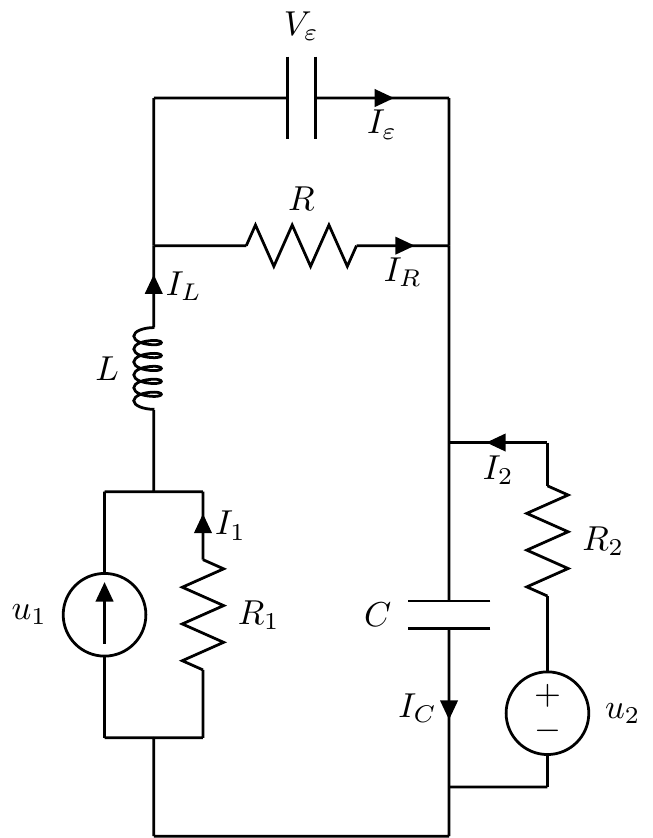}}};
\node at (4.5,-5) {(c)};

\end{tikzpicture}
  \caption{(a) Electric circuit $\Sigma_1$ with a nonlinear resistive load. (b) The regularization of circuit (a), denoted by $\Sigma_2$. (c) The controlled circuit $\Sigma_3$. }
  \label{fc1}
\end{figure}

Let us consider the electric circuit $\Sigma_1$ as shown in Figure \ref{fc1}.(a), where the Capacitor $C$ and Inductor $L$ are usual elements, but the Resistor $R$ is assumed to be nonlinear, that is $I_R=f(V_R)$ where $f$ is some nonlinear function and $I_R$ and $V_R$ denote the current and voltage of $R$ respectively, see Example 5. of \cite{smale1972mathematical}.

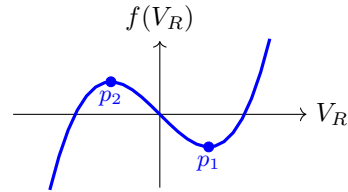
\begin{figure}[htbp]\centering
\begin{tikzpicture}[scale=0.65]
\draw[->] (-3,0) -- (3,0) node[right] {$V_R$};
\draw[->] (0,-1.5) -- (0,1.5) node[above] {$f(V_R)$};
\draw[domain=-2.25:2.25,very thick,blue] plot (\x,0.33333*\x*\x*\x-\x);
\draw[blue,fill=blue] (1,-2/3) circle (0.1) node[below]{\small $p_1$};
\draw[blue,fill=blue] (-1,2/3) circle (0.1) node[below]{\small $p_2$};
\end{tikzpicture}
\caption{The characteristic curve of the nonlinear Resistor $R$.}
\label{f:r}
\end{figure}

For the purpose of this example we shall assume that
\begin{equation}
f(V_R)=\frac{1}{3}V_R^3-V_R,
\end{equation}
and then the characteristic curve of the nonlinear Resistor $R$ is as depicted in Figure \ref{f:r}. We remark that the chosen type of nonlinear behavior is qualitatively the same as the one in \cite{smale1972mathematical}, and also appears in several other nonlinear elements, such as tunnel diodes \cite{Reissig1}.

The characteristic curve depicted in Figure \ref{f:r} has two fold points $p_{1,2}$ located at $(V_R,I_R)=(\pm 1,\mp \frac{2}{3})$. At these points the differential equation describing the behavior of the circuit becomes singular. To overcome this it is proposed to regularize $\Sigma_1$ by adding a parasitic capacitance in parallel to $R$ as depicted in Figure \ref{fc1}.(b) (electric circuit $\Sigma_2$), see Example 6 of \cite{smale1972mathematical} for more details, and \cite{ihrig1975regularization}. The capacitance $C_\ve$ is assumed to be small, e.g. $C_\ve=\ve$. Thus, as $\ve\to 0$, the behavior of of  circuit $\Sigma_2$ approaches that of $\Sigma_1$. Let $(x_1,x_2,z)=(I_L,V_C,V_{\ve})$, where $V_{\ve}$ denotes the voltage at the capacitor $C_\ve$. The equations describing the behavior of the circuit $\Sigma_2$ read as
\begin{equation}\label{eq:ex1}
\begin{split}
L\dot x_1 &= -z-x_2\\
C\dot x_2 &= x_1\\
\ve\dot z &= -\frac{1}{3}z^3+z+x_1.
\end{split}
\end{equation}

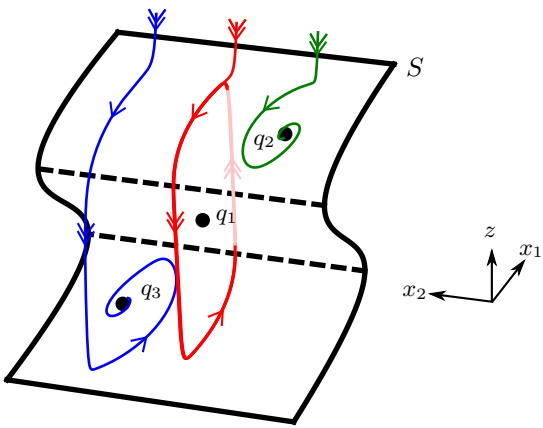
\begin{figure}[htbp]\centering
\begin{tikzpicture}
\pgftext{\includegraphics[scale=1]{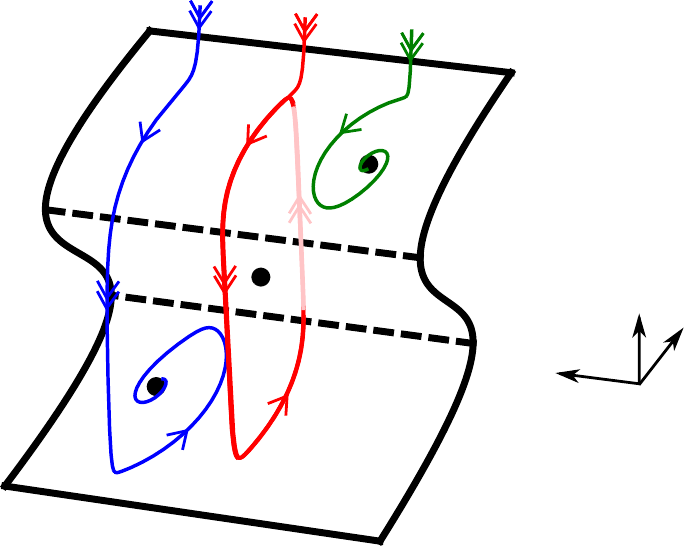}}
\node at (2,-1) {\small $x_2$};
\node at (3,-0.2) {\small $z$};
\node at (3.55,-0.45) {\small $x_1$};
\node at (2,2) {$S$};
\node at (0,1) {\small$q_2$};
\node at (-.5,0) {\small$q_1$};
\node at (-1.5,-1) {\small$q_3$};
\end{tikzpicture}
\caption{Trajectories of the open-loop system \eqref{eq:ex1}. The surface is the critical manifold $S$ while dashed lines represent lines of folds, and the dots stand for the equilibrium points. Note the existence of a limit cycle, which can be conjectured from the stability of $q_2$ and $q_3$. However such analysis falls off the scope of this document and shall not be discussed further.}
\label{f:ex2}
\end{figure}

We immediately note that the corresponding critical manifold is actually given by the characteristic equation of the nonlinear resistor, namely
\begin{equation}
S=\left\{ (x_1,x_2,z)\in\R^3\,|\, x_1=\frac{1}{3}z^3-z \right\}.
\end{equation}
Moreover, it is straightforward to see that the region of $S$ between $p_1$ and $p_2$ is repelling while the rest of $S$ is attracting. Furthermore, one can easily show that \eqref{eq:ex1} has three equilibrium points
$\left\{q_1,q_2,q_3\right\}=\left\{ (0,0,0),(0,-\sqrt{3},\sqrt{3}), (0,\sqrt{3},-\sqrt{3}) \right\}$, where $q_1$ is unstable, while $q_2$ and $q_3$ are stable. With this information we can qualitatively describe the dynamics of \eqref{eq:ex1} as follows: for initial conditions away from $S$, the trajectories of \eqref{eq:ex1} quickly approach a stable region of $S$, and then evolve around it. Here two things may happen, trajectories may converge to an equilibrium point $q_2$ or $q_3$, or they can approach a fold point. When a trajectory reaches a fold point, then the trajectory jumps towards a stable region of $S$ and then follows the same behavior as as described before. A sample of trajectories of \eqref{eq:ex1} is provided in Figure \ref{f:ex2}.

Our goal, however, is to set as operating point a fold point. In order to do this we introduce controllers as depicted in Figure \ref{fc1}.(c) leading to circuit $\Sigma_3$, where we add a current source $u_1$ in parallel with the resistor $R_1$, all these in series with the inductor $L$. The capacitor voltage $x_2$ is modulated by placing a voltage source $u_2$ and a resistor $R_2$ in series, all these in parallel with the capacitor $C$.  Both "virtual" resistors $R_1$ and $R_2$ are understood to be part of the controller, and in physical terms are for proper coupling of the inputs and may also serve as tuning parameters in a real implementation. 


The behavior of the circuit $\Sigma_3$ is described by
\begin{equation}\label{eq:ex2}
\begin{split}
L\dot x_1 &= -z-x_2 -R_1x_1+R_1u_1\\
C\dot x_2 &= x_1+\frac{x_2}{R_2}-\frac{u_2}{R_2}\\
\ve\dot z &= -\frac{1}{3}z^3+z+x_1.
\end{split}
\end{equation}
Let us choose, for example, the operating point $P=(x_1,x_2,z)=(-\frac{2}{3},0,1)$. We remark that since $P$ is a non-hyperbolic point, the classical theory e.g. \cite{Kokotovic:1986:SPM:576779} cannot be used, and instead we shall employ the theory developed in this paper. The first step is to define new coordinates $(X_1,X_2,Z)=(-x_1-\frac{2}{3},x_2,z-1)$, in this way \eqref{eq:ex2} reads as
\begin{equation}\label{eq:ex3}
\begin{split}
L\dot X_1 &= 1-\frac{2}{3}R_1 - R_1X_1 + X_2 + Z - R_1u_1\\
C\dot X_2 &= -\frac{2}{3} - X_1 + \frac{X_2}{R_2}-\frac{u_2}{R_2}\\
\ve\dot Z &= -(Z^2+X_1)-\frac{1}{3}Z^3.
\end{split}
\end{equation}
Note that up to leading order terms, \eqref{eq:ex3} is of the form studied above. After the change of coordinates the operating point is located at the origin of \eqref{eq:ex3}. According to Theorem \ref{thm:mainAda} the controller that stabilizes $P$ is given by
\begin{equation}
\begin{split}
u_1 &=-\frac{L}{R_1}\left( -\frac{1}{L} +\frac{2}{3L}R_1 -\frac{c_1}{\ve^{2/3}}X_1+\frac{c_0c_1+\gamma}{\ve^{1/3}}Z+\beta \right)     \\
u_2 &=-CR_2\left( \frac{2}{3C}-\frac{c_1}{\ve^{2/3}}X_2\right),
\end{split}
\end{equation}
where $\beta=\ve^{-2/3}\left( -c_1 Z^2+(2\ve^{-1/3}Z-c_0)(Z^2+X_1) \right)$.

\begin{figure}[htbp]\centering
\includegraphics[]{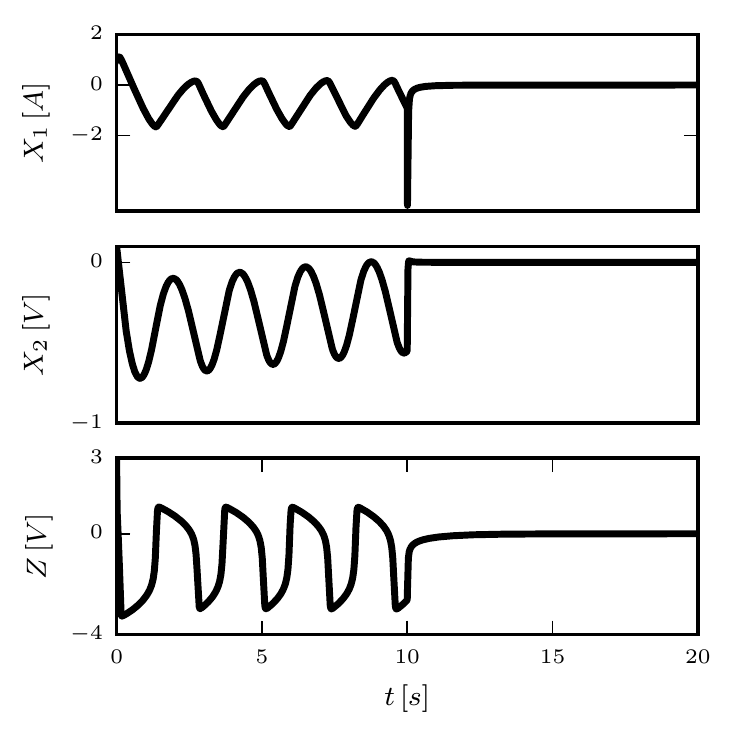}
\caption{Simulation results of the controlled circuit $\Sigma_3$ depicted in Figure \ref{f:ex2}.(c) and with the controller of Theorem \ref{thm:mainAda}.}
\label{f:res}
\end{figure}

To witness the effects of the controller, let us choose parameters: $C=1$F, $L=1$H, $R_1=R_2=1\Omega$, $\ve=0.05$, and controller gains $c_0=\gamma=1$ and $c_1=10$. Next we choose initial conditions near the limit cycle, to allow the system oscillate. For simulation purposes we let the system evolve 10 seconds in open-loop. Then, at $t=10$s we activate the controller. The results are shown in Figure \ref{f:res}, where we see that when the controller takes action, the trajectories quickly converge to the operating point $P$.

\section{Conclusions}\label{sec:conclusions}

In this paper we have introduced the geometric desingularization technique to control systems. The main contribution is a controller design method for non-hyperbolic fold points of slow-fast systems. With this we are able to deal with singular perturbation problems for which the classical theory does not apply. An essential feature of our contribution is that the controller only actuates on the slow variables, making it more suitable for applications. As a case study, we have provided a controller based on the backstepping algorithm that renders the origin of a SFCS globally asymptotically stable.

Further research directions in view of the potential applications include: the assumption that the slow system is under-actuated, output feedback control,  trajectory and path-following along sets of non-hyperbolic points, and the extension of the theory to a general class of SFCS with one fast variable but with arbitrarily degenerate non-hyperbolic points. The latter class of systems are of the form
\begin{equation}
\begin{split}
\dot x &= f(x,z,\ve) + B(x,z,\ve)u\\
\ve\dot z &= -\left( z^k + \sum_{i=1}^{k-1}x_iz^{i-1} \right) + O(\ve), 
\end{split}
\end{equation}
where $k\geq 2$.
Another challenging framework is to consider non-hyperbolic slow-fast systems over networks.




\bibliographystyle{plain}
\bibliography{Mendeley}

\end{document}